\frontpagetrue
\input tables
\def\PsfigVersion{1.9}
\ifx\undefined\psfig\else \fi

%

\let\LaTeXAtSign=\@
\let\@=\relax
\edef\psfigRestoreAt{\catcode`\@=\number\catcode`@\relax}
\catcode`\@=11\relax
\newwrite\@unused
\def\ps@typeout#1{{\let\protect\string\immediate\write\@unused{#1}}}
\ps@typeout{psfig/tex \PsfigVersion}


\def\figurepath{./}

%
%
\def\@nnil{\@nil}
\def\@empty{}
\def\@psdonoop#1\@@#2#3{}
\def\@psdo#1:=#2\do#3{\edef\@psdotmp{#2}\ifx\@psdotmp\@empty \else
    \expandafter\@psdoloop#2,\@nil,\@nil\@@#1{#3}\fi}
\def\@psdoloop#1,#2,#3\@@#4#5{\def#4{#1}\ifx #4\@nnil \else
       #5\def#4{#2}\ifx #4\@nnil \else#5\@ipsdoloop #3\@@#4{#5}\fi\fi}
\def\@ipsdoloop#1,#2\@@#3#4{\def#3{#1}\ifx #3\@nnil 
       \let\@nextwhile=\@psdonoop \else
      #4\relax\let\@nextwhile=\@ipsdoloop\fi\@nextwhile#2\@@#3{#4}}
\def\@tpsdo#1:=#2\do#3{\xdef\@psdotmp{#2}\ifx\@psdotmp\@empty \else
    \@tpsdoloop#2\@nil\@nil\@@#1{#3}\fi}
\def\@tpsdoloop#1#2\@@#3#4{\def#3{#1}\ifx #3\@nnil 
       \let\@nextwhile=\@psdonoop \else
      #4\relax\let\@nextwhile=\@tpsdoloop\fi\@nextwhile#2\@@#3{#4}}
%
\ifx\undefined\fbox
\newdimen\fboxrule
\newdimen\fboxsep
\newdimen\ps@tempdima
\newbox\ps@tempboxa
\fboxsep = 3pt
\fboxrule = .4pt
\long\def\fbox#1{\leavevmode\setbox\ps@tempboxa\hbox{#1}\ps@tempdima\fboxrule
    \advance\ps@tempdima \fboxsep \advance\ps@tempdima \dp\ps@tempboxa
   \hbox{\lower \ps@tempdima\hbox
  {\vbox{\hrule height \fboxrule
          \hbox{\vrule width \fboxrule \hskip\fboxsep
          \vbox{\vskip\fboxsep \box\ps@tempboxa\vskip\fboxsep}\hskip 
                 \fboxsep\vrule width \fboxrule}
                 \hrule height \fboxrule}}}}
\fi
%
%
\newread\ps@stream
\newif\ifnot@eof       
\newif\if@noisy        
\newif\if@atend        
\newif\if@psfile       
%
%
{\catcode`\%=12\global\gdef\epsf@start{
\def\epsf@PS{PS}
\def\epsf@getbb#1{%
%
%
\openin\ps@stream=#1
\ifeof\ps@stream\ps@typeout{Error, File #1 not found}\else
%
%
   {\not@eoftrue \chardef\other=12
    \def\do##1{\catcode`##1=\other}\dospecials \catcode`\ =10
    \loop
       \if@psfile
	  \read\ps@stream to \epsf@fileline
       \else{
	  \obeyspaces
          \read\ps@stream to \epsf@tmp\global\let\epsf@fileline\epsf@tmp}
       \fi
       \ifeof\ps@stream\not@eoffalse\else
%
%
       \if@psfile\else
       \expandafter\epsf@test\epsf@fileline:. \\%
       \fi
%
%
          \expandafter\epsf@aux\epsf@fileline:. \\%
       \fi
   \ifnot@eof\repeat
   }\closein\ps@stream\fi}%
%
%
\long\def\epsf@test#1#2#3:#4\\{\def\epsf@testit{#1#2}
			\ifx\epsf@testit\epsf@start\else
\ps@typeout{Warning! File does not start with `\epsf@start'.  It may not be a PostScript file.}
			\fi
			\@psfiletrue} 
%
%
{\catcode`\%=12\global\let\epsf@percent=
%
%
%
\long\def\epsf@aux#1#2:#3\\{\ifx#1\epsf@percent
   \def\epsf@testit{#2}\ifx\epsf@testit\epsf@bblit
	\@atendfalse
        \epsf@atend #3 . \\%
	\if@atend	
	   \if@verbose{
		\ps@typeout{psfig: found `(atend)'; continuing search}
	   }\fi
        \else
        \epsf@grab #3 . . . \\%
        \not@eoffalse
        \global\no@bbfalse
        \fi
   \fi\fi}%
%
%
\def\epsf@grab #1 #2 #3 #4 #5\\{%
   \global\def\epsf@llx{#1}\ifx\epsf@llx\empty
      \epsf@grab #2 #3 #4 #5 .\\\else
   \global\def\epsf@lly{#2}%
   \global\def\epsf@urx{#3}\global\def\epsf@ury{#4}\fi}%
%
%
\def\epsf@atendlit{(atend)} 
\def\epsf@atend #1 #2 #3\\{%
   \def\epsf@tmp{#1}\ifx\epsf@tmp\empty
      \epsf@atend #2 #3 .\\\else
   \ifx\epsf@tmp\epsf@atendlit\@atendtrue\fi\fi}


\chardef\psletter = 11 
\chardef\other = 12

\newif \ifdebug 
\newif\ifc@mpute 
\c@mputetrue 

\let\then = \relax
\def\r@dian{pt }
\let\r@dians = \r@dian
\let\dimensionless@nit = \r@dian
\let\dimensionless@nits = \dimensionless@nit
\def\internal@nit{sp }
\let\internal@nits = \internal@nit
\newif\ifstillc@nverging
\def \Mess@ge #1{\ifdebug \then \message {#1} \fi}

{ 
	\catcode `\@ = \psletter
	\gdef \nodimen {\expandafter \n@dimen \the \dimen}
	\gdef \term #1 #2 #3%
	       {\edef \t@ {\the #1}
		\edef \t@@ {\expandafter \n@dimen \the #2\r@dian}%
		\t@rm {\t@} {\t@@} {#3}%
	       }
	\gdef \t@rm #1 #2 #3%
	       {{%
		\count 0 = 0
		\dimen 0 = 1 \dimensionless@nit
		\dimen 2 = #2\relax
		\Mess@ge {Calculating term #1 of \nodimen 2}%
		\loop
		\ifnum	\count 0 < #1
		\then	\advance \count 0 by 1
			\Mess@ge {Iteration \the \count 0 \space}%
			\Multiply \dimen 0 by {\dimen 2}%
			\Mess@ge {After multiplication, term = \nodimen 0}%
			\Divide \dimen 0 by {\count 0}%
			\Mess@ge {After division, term = \nodimen 0}%
		\repeat
		\Mess@ge {Final value for term #1 of 
				\nodimen 2 \space is \nodimen 0}%
		\xdef \Term {#3 = \nodimen 0 \r@dians}%
		\aftergroup \Term
	       }}
	\catcode `\p = \other
	\catcode `\t = \other
	\gdef \n@dimen #1pt{#1} 
}

\def \Divide #1by #2{\divide #1 by #2} 

\def \Multiply #1by #2
       {{
	\count 0 = #1\relax
	\count 2 = #2\relax
	\count 4 = 65536
	\Mess@ge {Before scaling, count 0 = \the \count 0 \space and
			count 2 = \the \count 2}%
	\ifnum	\count 0 > 32767 
	\then	\divide \count 0 by 4
		\divide \count 4 by 4
	\else	\ifnum	\count 0 < -32767
		\then	\divide \count 0 by 4
			\divide \count 4 by 4
		\else
		\fi
	\fi
	\ifnum	\count 2 > 32767 
	\then	\divide \count 2 by 4
		\divide \count 4 by 4
	\else	\ifnum	\count 2 < -32767
		\then	\divide \count 2 by 4
			\divide \count 4 by 4
		\else
		\fi
	\fi
	\multiply \count 0 by \count 2
	\divide \count 0 by \count 4
	\xdef \product {#1 = \the \count 0 \internal@nits}%
	\aftergroup \product
       }}

\def\r@duce{\ifdim\dimen0 > 90\r@dian \then   
		\multiply\dimen0 by -1
		\advance\dimen0 by 180\r@dian
		\r@duce
	    \else \ifdim\dimen0 < -90\r@dian \then  
		\advance\dimen0 by 360\r@dian
		\r@duce
		\fi
	    \fi}

\def\Sine#1%
       {{%
	\dimen 0 = #1 \r@dian
	\r@duce
	\ifdim\dimen0 = -90\r@dian \then
	   \dimen4 = -1\r@dian
	   \c@mputefalse
	\fi
	\ifdim\dimen0 = 90\r@dian \then
	   \dimen4 = 1\r@dian
	   \c@mputefalse
	\fi
	\ifdim\dimen0 = 0\r@dian \then
	   \dimen4 = 0\r@dian
	   \c@mputefalse
	\fi
	\ifc@mpute \then
		\divide\dimen0 by 180
		\dimen0=3.141592654\dimen0
		\dimen 2 = 3.1415926535897963\r@dian 
		\divide\dimen 2 by 2 
		\Mess@ge {Sin: calculating Sin of \nodimen 0}%
		\count 0 = 1 
		\dimen 2 = 1 \r@dian 
		\dimen 4 = 0 \r@dian 
		\loop
			\ifnum	\dimen 2 = 0 
			\then	\stillc@nvergingfalse 
			\else	\stillc@nvergingtrue
			\fi
			\ifstillc@nverging 
			\then	\term {\count 0} {\dimen 0} {\dimen 2}%
				\advance \count 0 by 2
				\count 2 = \count 0
				\divide \count 2 by 2
				\ifodd	\count 2 
				\then	\advance \dimen 4 by \dimen 2
				\else	\advance \dimen 4 by -\dimen 2
				\fi
		\repeat
	\fi		
			\xdef \sine {\nodimen 4}%
       }}

\def\Cosine#1{\ifx\sine\UnDefined\edef\Savesine{\relax}\else
		             \edef\Savesine{\sine}\fi
	{\dimen0=#1\r@dian\advance\dimen0 by 90\r@dian
	 \Sine{\nodimen 0}
	 \xdef\cosine{\sine}
	 \xdef\sine{\Savesine}}}	      

\def\psdraft{
	\def\@psdraft{0}
}
\def\psfull{
	\def\@psdraft{100}
}

\psfull

\newif\if@scalefirst
\def\psscalefirst{\@scalefirsttrue}
\def\psrotatefirst{\@scalefirstfalse}
\psrotatefirst

\newif\if@draftbox
\def\psnodraftbox{
	\@draftboxfalse
}
\def\psdraftbox{
	\@draftboxtrue
}
\@draftboxtrue

\newif\if@prologfile
\newif\if@postlogfile
\def\pssilent{
	\@noisyfalse
}
\def\psnoisy{
	\@noisytrue
}
\psnoisy
\newif\if@bbllx
\newif\if@bblly
\newif\if@bburx
\newif\if@bbury
\newif\if@height
\newif\if@width
\newif\if@rheight
\newif\if@rwidth
\newif\if@angle
\newif\if@clip
\newif\if@verbose
\def\@p@@sclip#1{\@cliptrue}

\newif\if@decmpr


\def\@p@@sfigure#1{\def\@p@sfile{null}\def\@p@sbbfile{null}
	        \openin1=#1.bb
		\ifeof1\closein1
	        	\openin1=\figurepath#1.bb
			\ifeof1\closein1
			        \openin1=#1
				\ifeof1\closein1%
				       \openin1=\figurepath#1
					\ifeof1
					   \ps@typeout{Error, File #1 not found}
						\if@bbllx\if@bblly
				   		\if@bburx\if@bbury
			      				\def\@p@sfile{#1}%
			      				\def\@p@sbbfile{#1}%
							\@decmprfalse
				  	   	\fi\fi\fi\fi
					\else\closein1
				    		\def\@p@sfile{\figurepath#1}%
				    		\def\@p@sbbfile{\figurepath#1}%
						\@decmprfalse
	                       		\fi%
			 	\else\closein1%
					\def\@p@sfile{#1}
					\def\@p@sbbfile{#1}
					\@decmprfalse
			 	\fi
			\else
				\def\@p@sfile{\figurepath#1}
				\def\@p@sbbfile{\figurepath#1.bb}
				\@decmprtrue
			\fi
		\else
			\def\@p@sfile{#1}
			\def\@p@sbbfile{#1.bb}
			\@decmprtrue
		\fi}

\def\@p@@sfile#1{\@p@@sfigure{#1}}

\def\@p@@sbbllx#1{
		\@bbllxtrue
		\dimen100=#1
		\edef\@p@sbbllx{\number\dimen100}
}
\def\@p@@sbblly#1{
		\@bbllytrue
		\dimen100=#1
		\edef\@p@sbblly{\number\dimen100}
}
\def\@p@@sbburx#1{
		\@bburxtrue
		\dimen100=#1
		\edef\@p@sbburx{\number\dimen100}
}
\def\@p@@sbbury#1{
		\@bburytrue
		\dimen100=#1
		\edef\@p@sbbury{\number\dimen100}
}
\def\@p@@sheight#1{
		\@heighttrue
		\dimen100=#1
   		\edef\@p@sheight{\number\dimen100}
}
\def\@p@@swidth#1{
		\@widthtrue
		\dimen100=#1
		\edef\@p@swidth{\number\dimen100}
}
\def\@p@@srheight#1{
		\@rheighttrue
		\dimen100=#1
		\edef\@p@srheight{\number\dimen100}
}
\def\@p@@srwidth#1{
		\@rwidthtrue
		\dimen100=#1
		\edef\@p@srwidth{\number\dimen100}
}
\def\@p@@sangle#1{
		\@angletrue
		\edef\@p@sangle{#1} 
}
\def\@p@@ssilent#1{ 
		\@verbosefalse
}
\def\@p@@sprolog#1{\@prologfiletrue\def\@prologfileval{#1}}
\def\@p@@spostlog#1{\@postlogfiletrue\def\@postlogfileval{#1}}
\def\@cs@name#1{\csname #1\endcsname}
\def\@setparms#1=#2,{\@cs@name{@p@@s#1}{#2}}
%
%
\def\ps@init@parms{
		\@bbllxfalse \@bbllyfalse
		\@bburxfalse \@bburyfalse
		\@heightfalse \@widthfalse
		\@rheightfalse \@rwidthfalse
		\def\@p@sbbllx{}\def\@p@sbblly{}
		\def\@p@sbburx{}\def\@p@sbbury{}
		\def\@p@sheight{}\def\@p@swidth{}
		\def\@p@srheight{}\def\@p@srwidth{}
		\def\@p@sangle{0}
		\def\@p@sfile{} \def\@p@sbbfile{}
		\def\@p@scost{10}
		\def\@sc{}
		\@prologfilefalse
		\@postlogfilefalse
		\@clipfalse
		\if@noisy
			\@verbosetrue
		\else
			\@verbosefalse
		\fi
}
%
%
\def\parse@ps@parms#1{
	 	\@psdo\@psfiga:=#1\do
		   {\expandafter\@setparms\@psfiga,}}
%
%
\newif\ifno@bb
\def\bb@missing{
	\if@verbose{
		\ps@typeout{psfig: searching \@p@sbbfile \space  for bounding box}
	}\fi
	\no@bbtrue
	\epsf@getbb{\@p@sbbfile}
        \ifno@bb \else \bb@cull\epsf@llx\epsf@lly\epsf@urx\epsf@ury\fi
}	
\def\bb@cull#1#2#3#4{
	\dimen100=#1 bp\edef\@p@sbbllx{\number\dimen100}
	\dimen100=#2 bp\edef\@p@sbblly{\number\dimen100}
	\dimen100=#3 bp\edef\@p@sbburx{\number\dimen100}
	\dimen100=#4 bp\edef\@p@sbbury{\number\dimen100}
	\no@bbfalse
}
\newdimen\p@intvaluex
\newdimen\p@intvaluey
\def\rotate@#1#2{{\dimen0=#1 sp\dimen1=#2 sp
		  \global\p@intvaluex=\cosine\dimen0
		  \dimen3=\sine\dimen1
		  \global\advance\p@intvaluex by -\dimen3
		  \global\p@intvaluey=\sine\dimen0
		  \dimen3=\cosine\dimen1
		  \global\advance\p@intvaluey by \dimen3
		  }}
\def\compute@bb{
		\no@bbfalse
		\if@bbllx \else \no@bbtrue \fi
		\if@bblly \else \no@bbtrue \fi
		\if@bburx \else \no@bbtrue \fi
		\if@bbury \else \no@bbtrue \fi
		\ifno@bb \bb@missing \fi
		\ifno@bb \ps@typeout{FATAL ERROR: no bb supplied or found}
			\no-bb-error
		\fi
		%
%
		\count203=\@p@sbburx
		\count204=\@p@sbbury
		\advance\count203 by -\@p@sbbllx
		\advance\count204 by -\@p@sbblly
		\edef\ps@bbw{\number\count203}
		\edef\ps@bbh{\number\count204}
		\if@angle 
			\Sine{\@p@sangle}\Cosine{\@p@sangle}
	        	{\dimen100=\maxdimen\xdef\r@p@sbbllx{\number\dimen100}
					    \xdef\r@p@sbblly{\number\dimen100}
			                    \xdef\r@p@sbburx{-\number\dimen100}
					    \xdef\r@p@sbbury{-\number\dimen100}}
%
                        \def\minmaxtest{
			   \ifnum\number\p@intvaluex<\r@p@sbbllx
			      \xdef\r@p@sbbllx{\number\p@intvaluex}\fi
			   \ifnum\number\p@intvaluex>\r@p@sbburx
			      \xdef\r@p@sbburx{\number\p@intvaluex}\fi
			   \ifnum\number\p@intvaluey<\r@p@sbblly
			      \xdef\r@p@sbblly{\number\p@intvaluey}\fi
			   \ifnum\number\p@intvaluey>\r@p@sbbury
			      \xdef\r@p@sbbury{\number\p@intvaluey}\fi
			   }
			\rotate@{\@p@sbbllx}{\@p@sbblly}
			\minmaxtest
			\rotate@{\@p@sbbllx}{\@p@sbbury}
			\minmaxtest
			\rotate@{\@p@sbburx}{\@p@sbblly}
			\minmaxtest
			\rotate@{\@p@sbburx}{\@p@sbbury}
			\minmaxtest
			\edef\@p@sbbllx{\r@p@sbbllx}\edef\@p@sbblly{\r@p@sbblly}
			\edef\@p@sbburx{\r@p@sbburx}\edef\@p@sbbury{\r@p@sbbury}
		\fi
		\count203=\@p@sbburx
		\count204=\@p@sbbury
		\advance\count203 by -\@p@sbbllx
		\advance\count204 by -\@p@sbblly
		\edef\@bbw{\number\count203}
		\edef\@bbh{\number\count204}
}
%
%
\def\in@hundreds#1#2#3{\count240=#2 \count241=#3
		     \count100=\count240	
		     \divide\count100 by \count241
		     \count101=\count100
		     \multiply\count101 by \count241
		     \advance\count240 by -\count101
		     \multiply\count240 by 10
		     \count101=\count240	
		     \divide\count101 by \count241
		     \count102=\count101
		     \multiply\count102 by \count241
		     \advance\count240 by -\count102
		     \multiply\count240 by 10
		     \count102=\count240	
		     \divide\count102 by \count241
		     \count200=#1\count205=0
		     \count201=\count200
			\multiply\count201 by \count100
		 	\advance\count205 by \count201
		     \count201=\count200
			\divide\count201 by 10
			\multiply\count201 by \count101
			\advance\count205 by \count201
		     \count201=\count200
			\divide\count201 by 100
			\multiply\count201 by \count102
			\advance\count205 by \count201
		     \edef\@result{\number\count205}
}
\def\compute@wfromh{
		\in@hundreds{\@p@sheight}{\@bbw}{\@bbh}
		\edef\@p@swidth{\@result}
}
\def\compute@hfromw{
	        \in@hundreds{\@p@swidth}{\@bbh}{\@bbw}
		\edef\@p@sheight{\@result}
}
\def\compute@handw{
		\if@height 
			\if@width
			\else
				\compute@wfromh
			\fi
		\else 
			\if@width
				\compute@hfromw
			\else
				\edef\@p@sheight{\@bbh}
				\edef\@p@swidth{\@bbw}
			\fi
		\fi
}
\def\compute@resv{
		\if@rheight \else \edef\@p@srheight{\@p@sheight} \fi
		\if@rwidth \else \edef\@p@srwidth{\@p@swidth} \fi
}
%
\def\compute@sizes{
	\compute@bb
	\if@scalefirst\if@angle
	\if@width
	   \in@hundreds{\@p@swidth}{\@bbw}{\ps@bbw}
	   \edef\@p@swidth{\@result}
	\fi
	\if@height
	   \in@hundreds{\@p@sheight}{\@bbh}{\ps@bbh}
	   \edef\@p@sheight{\@result}
	\fi
	\fi\fi
	\compute@handw
	\compute@resv}

%
%
\def\psfig#1{\vbox {
	%
	\ps@init@parms
	\parse@ps@parms{#1}
	\compute@sizes
	\ifnum\@p@scost<\@psdraft{
		\special{ps::[begin] 	\@p@swidth \space \@p@sheight \space
				\@p@sbbllx \space \@p@sbblly \space
				\@p@sbburx \space \@p@sbbury \space
				startTexFig \space }
		\if@angle
			\special {ps:: \@p@sangle \space rotate \space} 
		\fi
		\if@clip{
			\if@verbose{
				\ps@typeout{(clip)}
			}\fi
			\special{ps:: doclip \space }
		}\fi
		\if@prologfile
		    \special{ps: plotfile \@prologfileval \space } \fi
		\if@decmpr{
			\if@verbose{
				\ps@typeout{psfig: including \@p@sfile.Z \space }
			}\fi
			\special{ps: plotfile "`zcat \@p@sfile.Z" \space }
		}\else{
			\if@verbose{
				\ps@typeout{psfig: including \@p@sfile \space }
			}\fi
			\special{ps: plotfile \@p@sfile \space }
		}\fi
		\if@postlogfile
		    \special{ps: plotfile \@postlogfileval \space } \fi
		\special{ps::[end] endTexFig \space }
		\vbox to \@p@srheight sp{
			\hbox to \@p@srwidth sp{
				\hss
			}
		\vss
		}
	}\else{
		\if@draftbox{		
			\hbox{\frame{\vbox to \@p@srheight sp{
			\vss
			\hbox to \@p@srwidth sp{ \hss \@p@sfile \hss }
			\vss
			}}}
		}\else{
			\vbox to \@p@srheight sp{
			\vss
			\hbox to \@p@srwidth sp{\hss}
			\vss
			}
		}\fi

	}\fi
}}
\psfigRestoreAt
\let\@=\LaTeXAtSign

\def \bkl{\hfil\break}
\hoffset=1cm
\PHYSREV
\voffset=1cm
\let\hf=\hfill
\def\9{\hphantom0}
\def\ep{e$^+$e$^-\;$}
\def\epq{e$^+$e$^- \rightarrow q {\overline q}\;$}
\def\epw{e$^+$e$^- \rightarrow W^+W^-\;$}
\def\epz{e$^+$e$^- \rightarrow Z^0Z^0\;$}
\def\epb{e$^+$e$^- \rightarrow b {\overline b}\;$}
\def\epqt{e$^+$e$^- \rightarrow q {\overline q}\; (q \ne t)\;$}
\def\ept{e$^+$e$^- \rightarrow t {\overline t}\;$}
\def\als{$\alpha_S$}
 
 \hsize=6in
   \vsize=8.5in
 \singlespace

\bigskip
	{\line{ 	\hfill\vbox{
	\hbox{SCIPP 96/45}\hbox{hep-ex/9612013}\hbox{November 1996}
		\vskip1in 	}}}

\title{IDENTIFICATION OF A SAMPLE OF
\epq
EVENTS FOR THE PRECISE MEASUREMENT OF $\alpha_S$
IN HIGH ENERGY ELECTRON--POSITRON COLLIDERS}
 
\vskip1cm
 
\centerline{Bruce A. Schumm}
\centerline{\em University of California, Santa Cruz}
\centerline{\em Santa Cruz, CA 95064}
\centerline{\em schumm@scipp.ucsc.edu}
 \vfill
 
\centerline{\bf Abstract}
 
\noindent
Based on the PYTHIA physics simulation package, and a fast
simulation of the proposed detector for the Next Linear
Collider, a set of cuts is identified which
leads to a sample of \epq events appropriate for the precise
measurement of \als\ in \ep annihilation at $\sqrt{s} = 500$
GeV/c$^2$. Using these cuts,
the systematic uncertainty on \als\ associated
with correcting for selection cut biases and remaining
non-\epqt contamination is expected to be less than
$\pm 1\%$.
This work was done as part of a study of the
prospects for the precise measurement of \als\ at future
High Energy Physics facilities, undertaken
for the
1996 Snowmass Workshop on New Directions in High Energy
Physics.
 
\vfill
\eject

\noindent
{\bf Introduction}
 
The approach of measuring \als\ via the reconstruction of
hadronic observables in the process \epqt at high energy
is thought to be an ideal place to perform a high $Q^2$
measurement of \als. With the calculation
of these observables done to next-to-next-to-leading
order in perturbative QCD -- a very likely prospect
on the time scale of data taking at the NLC -- it is
expected that theoretical uncertainties associated
with the truncation of the perturbative series, as well
as uncalculated non-perturbative effects, will be on the
order of $\pm 1\%$ [1].
Thus, it is important that expectations for the size of
experimental uncertainties
associated with the measurement of \als\ in high energy
\ep collisions not exceed $\pm 1\%$.
This paper discusses the results of a study in which event
cuts were identified for which
the systematic uncertainty on \als\ associated
with correcting for selection cut biases and remaining
non-\epqt contamination is expected to be less than
$\pm 1\%$. This work was done as part of a study of the
prospects for the precise measurement of \als\ at future
High Energy Physics facilities, undertaken
for the
1996 Snowmass Workshop on New Directions in High Energy
Physics.
 
\bigskip\noindent
{\bf The European Linear Collider QCD Working Group Event Selection}
 
At $\sqrt{s} = 500$GeV in \ep collisions, the Born-level
event rate is dominated by annihilation to $W^+W^-$
pairs (7.0 pb). Annihilation to light quark (udscb) pairs,
the process of interest for most approaches to
measuring \als\ at a high energy linear collider, has
a cross section smaller by a over factor of two (3.1 pb).
Annihilation to top pairs (0.3 pb) and $Z^0Z^0$ pairs
(0.4 pb) also form backgrounds which need to be suppressed
when identifying a sample of events for QCD analyses.
 
In a study performed by the European Working Group on QCD at
a High Energy Linear \ep Collider [2], a set of cuts was identified
which result in an 83\% pure sample of \epqt events. These
cuts are presented in Table 1, which is reproduced from
Reference [2]. In particular, a `hemisphere mass' cut, requiring
that at least one of the two thrust hemispheres
have an invariant
mass less than 13\% of the total visible energy in the event,
was included to suppress \ept and boson pair events.
After the application of the cuts, backgrounds to
\epqt in the event sample are dominated by
W-pair (11\%) and \ept (6\%) events.
This sample is appropriate for many
QCD analyses in \ep annihilation at $\sqrt{s} = 500$GeV.
 
\pageinsert
\vbox{  
   \tenpoint \baselineskip=12pt   \narrower
 \noindent
{\bf Table 1.}\enskip 
European Working Group event selection cuts, reproduced
 from Reference [2]. The NLC Working Group cuts, which are
 the subject of this paper, are identical to these cuts up
 to  the replacement of the hemisphere mass cut with a heavy
 quark anti-tag, and the removal of all events produced with
 nominally left-handed electron beam polarization.
}  \vskip1pc
\tablewidth=13cm
\thicksize=\thinsize
 \begintable
  Cut                                 |
  Value                             \cr
  Particle multiplicity\hf    | $ N_{charged}  \geq 8$ \nr
  Polar angle of thrust axis\hf |  $ \vert(\cos(\theta)_T)\vert <0.8$ \nr
  Visible energy  \hf     | ${E_{vis} \over E_{cm}} > 0.5$ \nrneg{19pt}
    |   \nr
  Longitudinal momentum balance\hf  | ${\left\vert\left(\sum{p_z}
           \right)\right\vert \over E_{vis}} < 0.4$  \nr
  Minimum hemisphere mass \hf | M$_1$ and M$_2 > 3$ GeV \nr
  Hemisphere multiplicity \hf  | $N^{1,2}_{charged} \geq 4$ \nr
  Hemisphere mass      \hf | $\left({M_1 \over E_{vis}}\ {\rm or}\
                       {M_2 \over E_{vis}}\right) < 0.13$ \nrneg{17pt}
    |
  \endtable
\vfill
   \tenpoint \baselineskip=12pt   \narrower
\centerline{\psfig{file=nlcqcd1.postscr,angle=90,height=8cm}}
\vskip6pt\noindent
{\bf Fig.~1.}\enskip
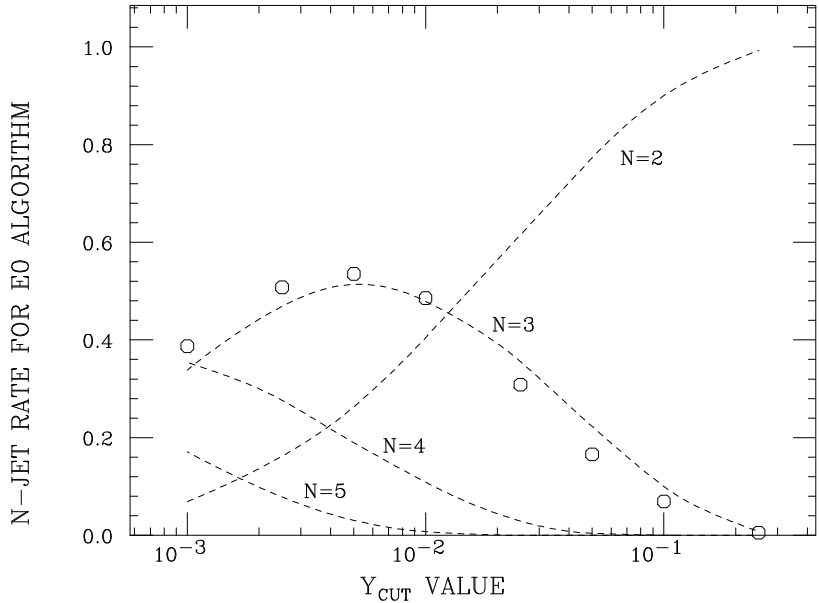
Comparison of E0 algorithm [3]
 three jet rate before (dashed lines)
 and after (open circles) the European Working
 Group event selection cuts, as a function of $y_{cut}$.
 To emphasize the effects of event selection bias,
 the effects of non-\epqt contamination, initial state
 radiation, and beamstrahlung are not included
 in the comparison.
\endinsert

The cuts, in particular the hemisphere mass cut, introduce
a moderate bias against hard gluon radiation which must
be corrected for after the sample is analyzed. For example,
Figure 1
show a Monte Carlo
comparison (to be described in more detail below)
of the $E0$ algorithm [3] three jet rate as a function of $y_{cut}$
for \epqt
events between a sample of events for which no selection
cuts have been applied, and a sample for which the
European Working Group cuts have been applied. To isolate
the effect of event selection bias, the effects of
initial state radiation and
of backgrounds
from $t {\overline t}$ and boson pair production
have not been included
in this plot. In the region $ 10^{-2} < y_{cut} < 10^{-1}$
relevant for the measurement of \als, biases of 10-20\%
in the three jet rate are observed, adequate for a measurement of \als
to $\sim \pm 5\%$.

\bigskip\noindent
{\bf The NLC QCD Working Group Event Selection}
 
Recently, considerable interest has been generated in the
possibility of a precise measurement of \als, for which the
uncertainty of the value of \als\ (at the scale
$Q^2 = M_Z^2$) would approach the level of $\pm 1\%$ [4].
This paper reports the results of a study in which the
European Working Group cuts were modified in order provide
a sample of \epqt events for which the uncertainty in the
corrections for event selection bias and background contamination
are consistent with this goal. This was achieved by
conducting a Monte Carlo study for which the hemisphere
mass cut was removed, replacing it with a cut on
events which exhibit heavy quark decay characteristics (to
suppress \ept backgrounds), and by
using events produced
by right-hand polarized electron beam only (to suppress
W-pair production). Otherwise, the event selection followed
precisely that developed by the European Working Group.
In this paper, this modified set of cuts will be referred to
as the `NLC Working Group' cuts.

\bigskip\noindent
{\bf Simulation Procedure}
 
The various \ep annihilation channels discussed above were
simulated with the PYTHIA 5.7 Monte Carlo [5].
Events were analyzed at the stable hadron level, with
detector simulation parameterized as discussed below.
The effects of local initial state radiation (that
component of ISR independent of beam currents and
densities) were included in the event simulation
via the MSTP(11) flag.
The complementary component of initial state radiation,
commonly referred to as `beamstrahlung', is not available
within Pythia, and thus was not simulated. However, the
Pythia simulation using local ISR only accurately
reproduced the European Working Group results, which
included a beamstrahlung simulation. This supports
the notion that the {\it relative} contributions
from the various \ep annihilation processes should
be roughly independent of the amount of beamstrahlung
radiation generated by the collisions.
 
A final state particle from the Pythia event generation was included
in the calorimeter analysis if it was stable and interacting,
had an $|\cos(\theta)|$ relative to the beam direction of
less than 0.97, and, if charged, a
transverse momentum relative to the beam direction of greater
than 300 MeV/c.
A charged track was included in the tracking analysis provided
it had an $|\cos(\theta)|$ of less than 0.9.
 
 \break
\bigskip\noindent
{\bf Lifetime Antitag}
 
In this study, events were not included in the QCD sample if they
showed substantial evidence of the presence of heavy quarks in the
final state. The primary purpose of this requirement was to
suppress \ept events from the final sample, particularly after the
removal of the hemisphere mass cut. This cut was also very
efficient in removing \epb events from the sample, even though
these latter events are suitable for the \als\ analysis.
Before the lifetime antitag,
\epb events
comprised approximately
15\% of all \epqt events.
 
Specifically, an event was considered to contain heavy quarks
if it produced four or more tracks with impact parameters
relative to the \ep collision point which
differed from 0 by $3\sigma$ or greater in either the
$r-\phi$ or $r-z$ view.
The impact parameter resolution assumed for the tracking system
was that of the proposed NLC detector, which is as follows [6]:
in the $r-\phi$ view, 2.6 $\mu m$ and 13.7 $\mu m$ for the
asymptotic and multiple scattering terms, respectively, and
in the $r-z$ view, 10 $\mu m$ and 30 $\mu m$ for the asymptotic
and multiple scattering terms. After applying
all of the European Working Group cuts except the hemisphere
mass cut, the application of this lifetime antitag
removed 95\% of the remaining \ept events, 19\% of the
remaining \epw events, and 47\% of the remaining
\epz events, while retaining 68\% of the remaining \epqt signal
events.
 
\midinsert
\vskip1pc
\vbox{\narrower   \singlespace
 \noindent
{\bf Table 2.}\enskip 
Expected left-right cross section asymmetry $A_{LR}$
 $\sqrt{s} = 500$ GeV for various \ep annihilation processes.
}
\vskip6pt
\thicksize=\thinsize
\tablewidth=8cm
 \begintable
 {\bf Process}   |   $A_{LR}$   \cr
 \epqt  \hf      |   \90.45    \nr
 \epw   \hf      |  $>$0.99   \nr
 \ept   \hf      |   \90.35    \nr
 \epz   \hf      |   \90.30
 \endtable
\endinsert
 
\medskip\noindent
{\bf Use of Right Handed Electron Beam to Suppress \epw
     Background}
 
Table 2 shows the left-right asymmetry
$$A_{LR} = {\sigma_L - \sigma_R  \over \sigma_L + \sigma_R}$$
expected at $\sqrt{s} = 500$ GeV
for the four \ep annihilation processes under consideration
in this study. Processes with $A_{LR} > 0$ are suppressed, with
a degree proportional to the magnitude of $A_{LR}$, for running
exclusively with right handed electron beam. Running exclusively
right handed
electron beam with a substantial polarization is thus an
effective way to suppress \epw events.
 
Figure 2 shows the resulting NLC Working Group event sample
as a function of electron beam polarization. For a beam
polarization of 80\%, currently available in the SLAC
LINAC, the resulting event sample is 82\% pure, with contaminations
of 13\% for \epw, 4.0\% for \epz, and 1.2\% for \ept events.
Research and development on high polarization cathodes
continues at SLAC. For a 90\% electron beam polarization, these
fractions become 87\%, 7.8\%, 4.3\%, and 1.3\%, respectively.
 
\topinsert
   \tenpoint \baselineskip=12pt   \narrower
\centerline{\psfig{file=nlcqcd2.postscr,angle=90,height=8cm}}
\vskip6pt\noindent
{\bf Fig.~2.}\enskip
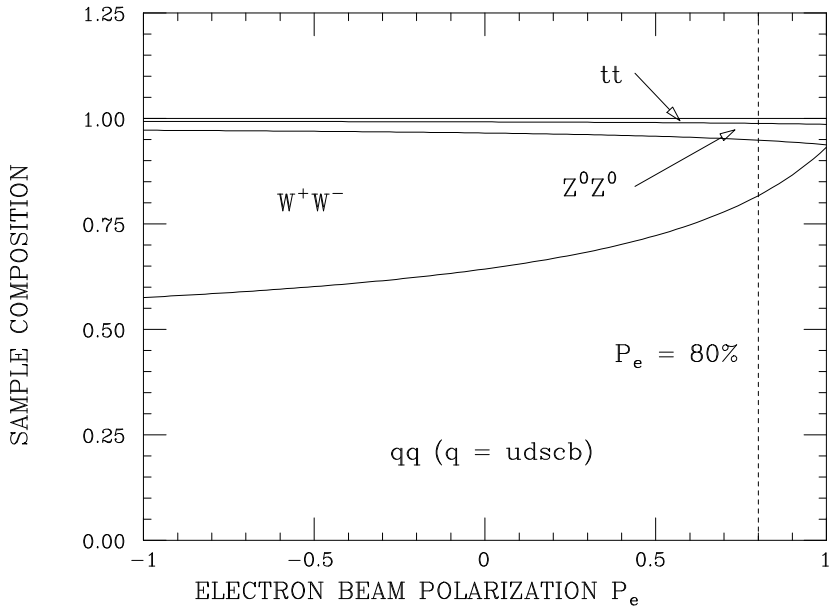
 Composition of the event sample resulting from the
 NLC Working Group event cuts, as a function of electron
 beam polarization.
\endinsert

\bigskip\noindent
{\bf Three Jet Rates for the NLC Working Group Sample}
 
Figure 3 shows the corresponding plot to Figure 1 for the NLC Working
Group event selection, isolating the effect of event selection
bias on hard gluon radiation. Differences between the unselected
sample three jet rate
and that of the sample after NLC Working
Group event selection (with the effects of
backgrounds and initial state
radiation excluded) are
at the few percent level. Thus, it is projected that the
uncertainty in correcting for the event selection bias
will be less than $\pm 1\%$.
 
\pageinsert
   \tenpoint \baselineskip=12pt   \narrower
\centerline{\psfig{file=nlcqcd3.postscr,angle=90,height=8cm}}
\vskip6pt\noindent
{\bf Fig.~3.}\enskip
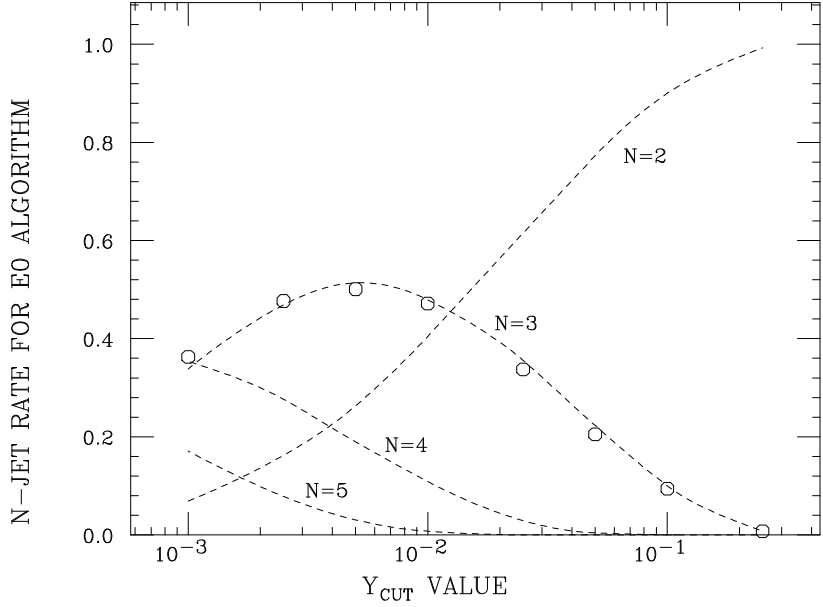
 Comparison of E0 algorithm [3]
 three jet rate before (dashed lines)
 and after (open circles) the NLC Working Group
 event selection cuts, as a function of $y_{cut}$.
 Again,
 to emphasize the effects of event selection bias,
 the effects of non-\epqt contamination, initial state
 radiation, and beamstrahlung are not included
 in the comparison.
\vskip1pc
\centerline{\psfig{file=nlcqcd4.postscr,angle=90,height=8cm}}
\vskip6pt\noindent
{\bf Fig.~4.}\enskip
 Comparison of the `raw' three jet rate
 (\epqt events at $\sqrt(s) = 500$GeV only; no event cuts
 or detector simulation), with the three jet rate
 expected for the NLC Working Group selection, including
 initial state radiation and all non-\epqt backgrounds.
\endinsert

Figure 4 is a plot similar to Figure 3, but with the effects
of ISR and event sample backgrounds included. In this and
all subsequent plots, an electron beam polarization
of 80\% is assumed.
Substantial
differences between the true three jet rate (dotted line)
and the expected experimental three jet rate (points) are
observed. As will be discussed immediately below, however, the
source of these differences are expected to be relatively
straightforward to model or measure from complementary data
samples, resulting in a correction to \als\ with an
uncertainty on the order of $\pm 1\%$.
 
The emission of a beamstrahlung or prompt radiated photon
prior to annihilation acts to lower the effective cms
energy of the \ep collision, and to give the annihilation
event a boost in the laboratory frame. With the European/NLC
working group cuts applied, the median fractional energy
loss due to initial state radiation is of order 50 GeV.
Both prompt radiation and beamstrahlung are
well-understood
physical processes, and can be modelled via QED and classical
electrodynamics. The uncertainty incurred by the correction
for these effects is expected to be small.
 
Figure 5 shows
a comparison of true and expected experimental three jet rates
after correcting for initial state radiation, assuming the
correction is known precisely. The remaining disagreement,
due in this plot to non-\epqt backgrounds and event selection
bias, is substantially smaller, particularly in the
region $10^{-2} < y_{cut} < 10^{-1}$.
 
\midinsert
   \tenpoint \baselineskip=12pt   \narrower
\centerline{\psfig{file=nlcqcd5.postscr,angle=90,height=8cm}}
\vskip6pt\noindent
{\bf Fig.~5.}\enskip
 Comparison of the `raw' three jet rate
 (\epqt events at $\sqrt(s) = 500$GeV only; no event cuts
 or detector simulation), with the three jet rate
 expected for the NLC Working Group selection, including
 non-\epqt backgrounds, but excluding the effects of
 initial state radiation.
\endinsert

Correcting for the $\sim 4\%$ \epz background should also be
systematically clean, given the wealth of precise data
on $Z^0 \rightarrow jets$ available from \ep annihilation
at the $Z^0$ pole. Figure 6 shows the three jet comparison
after correcting for \epz contamination, again assuming
the correction is known precisely.
 
Finally, Figure 7 shows the three jet comparison after correcting
for the $\sim 10\%$ \epw contamination. Depending on the magnitude
of the right-handed electron beam polarization, this is a
relative correction of 5-10\% on the three jet rate. In order
that the uncertainty on \als\ due to this correction be small compared to
$\pm 1\%$, it is necessary to know the magnitude of the
correction to 10-20\% of itself. Again, though, W boson decay
jet rates can be constrained directly with experimental data~--
for example, from jet rates observed opposite to
purely leptonic W decays in \epw events at high cms energy.

\pageinsert
   \tenpoint \baselineskip=12pt   \narrower
\centerline{\psfig{file=nlcqcd6.postscr,angle=90,height=8cm}}
\noindent
{\bf Fig.~6.}\enskip
 Comparison of the `raw' three jet rate
 (\epqt events at $\sqrt(s) = 500$GeV only; no event cuts
 or detector simulation), with the three jet rate
 expected for the NLC Working Group selection, including
 all non-\epqt backgrounds except \epz, and
 excluding the effects of
 initial state radiation.
\vskip2pc
\centerline{\psfig{file=nlcqcd7.postscr,angle=90,height=8cm}}
\noindent
{\bf Fig.~7.}\enskip
 Comparison of the `raw' three jet rate
 (\epqt events at $\sqrt(s) = 500$GeV only; no event cuts
 or detector simulation), with the three jet rate
 expected for the NLC Working Group selection, excluding all
 non-\epqt backgrounds except \ept, and
 excluding the effects of
 initial state radiation.
\endinsert

The remaining discrepancy observed in Figure 7, due to the effects
of \ept contamination and event selection bias, is not easily
constrained by data. On the other hand, the size of this
discrepancy is less than 5\% (relative) over most of the
range in $y_{cut}$ shown in the Figure. Thus, correcting
for these two final sources of discrepancy between the true
and observed three jet rate should also result in an uncertainty
on \als\ of less than 1\%.

\bigskip\noindent
{\bf Conclusion}
 
In the study reported in this paper, a set of event selection cuts
was identified which is expected to contribute a relative uncertainty
of no more than $\pm 1\%$ to a
measurement of
\als\ in high energy \ep annihilation.
This event selection is a modification to a set of cuts identified
earlier by the European Linear Collider QCD Working Group, with the
following changes: the hemisphere mass cut was removed, and
replaced by a heavy quark anti-tag, and all events produced with
(nominally) left-handed electron beam are discarded. Doing this
reduces difficult to constrain effects on the three jet rate
from event selection bias
and \ept contamination to manageable levels. Other backgrounds,
including \epw and \epz events remaining after the event selection cuts,
while somewhat substantially altering the three jet rate, can
be well constrained with existing or concurrent data, and thus
are not expected to contribute substantial systematic error
to the measurement of \als\ via three jet rates.
 
\bigskip\noindent
{\bf Acknowledgement}
 
The author would like to thank Richard Dubois of SLAC for his
substantial effort in getting PYTHIA to operate in a user-friendly
manner on the SLAC UNIX system, and for developing an
interface between PYTHIA output and the SLD analysis
stream.  This work is supported in part by the U.S. Department
of Energy, Contract \#DE-FG03-92ER40689.

 \vfill
 \eject
 
 \centerline{\bf References}
 
 \vskip .5truecm
 
 \point S. Kuhlman {\it et al.}, Physics and Technology of the
 Next Linear Collider,
 SLAC Report 485, June 1996, 105.
 
 \point S. Bethke, Proceedings of the Saariselka Workshop on
  Physics and Experiments With Linear Colliders, QCD161:W579:1991
  (1991) 575.
 
 \point S. Bethke {\it et al.}, Nucl. Phys. {\bf B370} (1992) 310.
 
 \point P. N. Burrows {\it et al.}, Prospects for the Precision
 Measurement of \als, to be submitted to the Proceedings of the
 1996 Workshop on New Directions in High Energy Physics,
 Snowmass, Colorado, June 1996, hep-ex/9612012.
 
 \point T. Sjostrand, Comput. Phys. Commun. {\bf 82} (1992) 74.
 
 \point R. Jacobsen, private communication.

 \vfill
 \eject

\bye